# Electronic properties of c-BN/diamond heterostructures for high-frequency high-power applications

Jeffrey T Mullen,[1] James A. Boulton,[1] Minghao Pan,[2] and Ki Wook Kim[1,2,*]

[1]*Department of Electrical and Computer Engineering,*
*North Carolina State University, Raleigh, NC 27695, USA*

[2]*Department of Physics, North Carolina State University, Raleigh, NC 27695, USA*

## Abstract

Using first principles calculations, this work investigates the suitability of diamond/c-BN heterojunctions for high frequency, high power device applications. The key quantities of band offsets and interface charge polarization are examined for different crystallographic orientations [(110), (111), or (100)], bond terminations (C-B or C-N), and substrates (diamond or c-BN). The results indicate that both the (111) and (100) structures with polar interfaces are likely to be a type-I alignment with the diamond conduction and valence band extrema nested within the c-BN bandgap, whereas the non-polar (110) counterpart may form type II as the valence band of c-BN is shifted down substantially lower. The valence band offsets are estimated to be around $0.2-0.55$ eV and $1.2-1.3$ eV for types I and II, respectively, with only a modest dependence on the order of layer stacking and bond termination. The (111) and (100) structures also show net charge polarization in a narrow region at the interface. The electron-deficient and electron-rich nature of the C-B and C-N bonding are found to induce charge redistribution leading to an essentially 2D sheet of negative and positive polarization, respectively, with a density on the order of $10^{12}-10^{13}$ $q/\text{cm}^2$ ($q = 1.6 \times 10^{-19}$ C). With the predicted band alignments suitable for carrier confinement as well as the possibility of the modulation and polarization doping, the diamond/c-BN heterostructures are a promising candidate for high-performance electronic devices with a highly conductive 2D channel. Both p-type and n-type devices appear possible with a judicious choice of the heterojunction configuration.

---

[*]Electronic address: `kwk@ncsu.edu`

## I. INTRODUCTION

Diamond and cubic boron nitride (c-BN) have been identified as two of the most promising candidates for high-frequency, high-power, and high-temperature applications. Their superior physical and chemical properties, including record thermal conductivity, ultra-wide bandgaps (UWBGs), extremely high breakdown fields, high electrical resistivity when undoped, high carrier mobility and saturation velocity, and negative electron affinity with efficient field emission are highlighted by characteristic figures of merit which surpass those of other competing materials such as SiC, GaN, AlN, and $\beta$-$Ga_2O_3$ [1–3]. Moreover, both diamond and c-BN are earth abundant and nontoxic. However, their promise as semiconductors that can deliver the ultimate performance and energy efficiency in a sustainable and environmentally friendly manner has not been fulfilled due to a number of technical roadblocks, one of which is the challenge in the doping for mobile carriers.

Like in any semiconductor-based platform, realization of UWBG devices requires efficient doping [4, 5]. Diamond can be doped with boron (p-type) during chemical vapor deposition (CVD) by using diborane, boron oxide or organoboron as the doping source [6]. However, the dopant concentrations achievable during CVD are usually low due to equilibrium thermodynamic solubility limits. On the other hand, p-type doping by ion implantation tends to create a large number of vacancy-interstitial pairs that must be annealed out to achieve significant electrical activation and carrier mobility [7]. Despite these difficulties, p-type doping has seen success in the past decade with a comparatively modest ionization energy ($\sim$0.37 eV for B at low concentrations). An acceptor density as high as $\sim 10^{21}$ cm$^{-3}$ has been reported in the literature via hot-filament CVD [8]. In comparison, n-type doping of diamond has presented more significant challenges [4, 7]. The n-type doping in diamond has been reported with substitutional N, P, and S dopants [4, 9] and interstitial Li and Na dopants [10, 11]. The results have been mixed despite decades of effort. While phosphorous doping via microwave plasma CVD (with the ionization energy of 0.57 eV) has reportedly reached a density as high as $\sim 8 \times 10^{19}$ cm$^{-3}$ along the $\langle 111 \rangle$ direction, going above dopant concentrations of $5 \times 10^{19}$ cm$^{-3}$ appears technically demanding in the (100) orientation [12]. These concentrations are still limited by thermodynamic solubility limits, and are substantially lower than those achieved with p-type dopants.

In the case of c-BN, group-II elements (such as Be, Mg, and Zn) substituted on the B sites should act as acceptors, while group-VI O and S at the N sites can be donors. Moreover, group-IV elements C and Si are amphoteric impurities and can substitute at both B and N sites. Experimentally, n-type conductivity in c-BN is generally achieved by S, Si, and C doping, whereas p-type is achieved by doping with Be, Mg, and Zn [5, 13, 14]. The activation energies of these dopants are mostly in the range of ∼0.2-0.3 eV. Both in-situ doping and ex-situ ion implantation have been demonstrated in c-BN bulk crystals and thin films with the dopant concentrations in excess of $10^{20}$ cm$^{-3}$ for n- and p-types [5, 14]. However, their key characteristics including the activation energies have shown a wide variation depending on the experimental conditions (e.g. bulk crystals vs. thin films; doping/growth techniques; dopant densities; etc.). It was also reported that the radiation damage caused by ion implantation at a high dose can lead to the loss of cubic phase and subsequent transformation to h-BN [15].

Nevertheless, a more fundamental challenge remains even when the issue of dopant control is resolved; i.e., the availability of mobile carriers due to the large dopant activation energies. With an activation energy of 0.25 eV (a relatively modest value in this material system), for instance, a rough estimate suggests far less than 1 % ionization at room temperature. The option of heavy doping and subsequent reduction in the ionization energy (e.g., see Ref. [16]) would lead to a drastic drop in the carrier mobility and drift velocity, rendering the material less than attractive in applications, where high-speed, high-energy-efficiency operation is desired. Indeed, the electron or hole mobility over 100 cm$^2$/V·s has rarely been report for both diamond and c-BN samples with a sizable carrier density ($\gtrsim$ mid $10^{17}$ cm$^{-3}$) except an outlier or two [17]. In fact, doping the channel itself is simply undesirable for high-energy-efficiency high-velocity transport even with dopants in a hydrogenic state [18]. As the ability to achieve heavy n- and p-doping is also essential for the formation of Ohmic contacts, the key to satisfying these often-conflicting requirements simultaneously is to induce the mobile carriers in the active device region without directly doping the channel.

A potential remedy may be to utilize the modulation and/or polarization doping concept through the application of diamond/c-BN heterojunctions [19], circumventing the limitations of high dopant activation energies without compromising their intrinsic advantages. Recent reports of epitaxial heterostructure synthesis in this nearly lattice-matched combination [20–23] provide another clear motivation to pursue the alternatives beyond those based on the bulk material systems. In the present work, we theoretically analyze the electronic properties of the diamond/c-BN heterostructures to examine the feasibility of this approach for high-frequency, high-power device applications. Our first-principle calculations clearly illustrate that the formation of high-mobility 2D carrier gas at the interface may indeed be achieved via the modulation and/or polarization doping with a proper selection of substrate (e.g., diamond vs. c-BN), crystallographic orientation [e.g., (111), (100), or (110) interfaces], and bond termination (e.g., C-B vs. C-N). Both n- and p-type devices appear possible. By comparison, the efforts based on the surface transfer doping via hydrogenation (i.e., hydrogen-terminated diamond) can only be applied to the p-type channels not to mention the observed significant degradation in the carrier mobility and the challenges on its long-term stability or durability [24].

## II. THEORETICAL MODEL

Density functional theory (DFT) as implemented in the QUANTUM ESPRESSO package with the Perdew-Burke-Ernzerhof functionals [25] is adopted for the investigation. The $k$-point sampling is performed on a $\Gamma$ centered Monkhorst-Pack grid. The bulk diamond and c-BN cells are optimized with an 8×8×8 grid and the density of states calculated with a denser 64×64×64 grid. The wave functions for all systems are expanded in a plane wave basis with kinetic energy cutoff of 260 Ry. Calculations with the pseudopotentials for the bulk materials show the diamond and c-BN lattice constants of 3.59 Å and 3.65 Å which compare well with experimental values of 3.567 Å [26] and 3.615 Å [27], respectively. Each of the heterostructures is configured in a slab geometry of the two materials in a supercell with sufficient vacuum space to prevent interactions in the periodic DFT calculations. Diamond and c-BN lattice constants differ by 1.6% indicating a small amount of strain between the two materials. Modelling the pseudomorphic heterostructure epitaxially achieved, we assume that the substrate (i.e., the thick bottom layer) remains unaffected while the relative thin overlayer grown on top is uniformly strained with the in-plane lattice constant matched to the bulk lattice constant of the substrate. As such, the geometry of this overlayer requires further optimization so that it also reflects the resultant adjustment in the growth direction following the Poisson's ratio [28]. For the calculations in the heterosystem, 16×16×1 grids are used.

Perhaps the most important quantity for characterizing the carrier confinement in hetero-junctions is the offset for valence and conduction bands. As the ab initio DFT calculations do not necessarily provide accurate bandgap calculations, we opt to determine the valence band offset (VBO) and use experimental values to find the corresponding value in the conduction band. The band offset is calculated from the equations [29, 30]

$$\Delta E_{\text{VBO}} = \left( E_V^{\text{sub}} - E_V^{\text{over}} \right)_{\text{bulk}} + \Delta U, \tag{1}$$

$$\Delta U = U_{\text{sub}} - U_{\text{over}}. \tag{2}$$

The first term in Eq. (1) denotes the difference in the valence band maxima of the two constituent bulk materials ($E_V^{\text{sub}}$ and $E_V^{\text{over}}$) that can be found from the respective density-of-states calculations. Note that in the case of $E_V^{\text{over}}$, the geometry of the overlayer deviates from its isotropic bulk form due to the pseudomorphic strain as described earlier. Accordingly, the degeneracy at the $\Gamma$ point of the Brillouin zone is lifted in the valence band. To avoid the subsequent ambiguity, an average of the split bands is used as the corresponding valence maximum. This average value for $E_V^{\text{over}}$ can be obtained from the self-consistent bulk characteristics of an effective cubic lattice (thus, isotropic) that is constructed from the strained overlayer by keeping its unit cell volume fixed [30].

The quantity $\Delta U$ is the alignment of the total electrostatic potentials on each side of the interface that accounts for the contributions of electronic and ionic charges in the heterojunction. In calculating this quantity, the oscillations of the electrostatic potential is removed with a suitable averaging scheme in order to focus on the differences at the junction. If the planar average parallel to the interface is represented by $U(z)$, then the macroscopic average is

$$\bar{U}(z) = \frac{1}{d} \int_{z-d/2}^{z+d/2} U(z')dz', \tag{3}$$

where $z$ is the direction perpendicular to the interface and $d$ is averaging window chosen to smooth the oscillations of the function [26, 29, 30]. Where the bulk characteristics of the material in each layer of the heterostructure are recovered, this averaging scheme gives a linear equation which can be extrapolated to the interface to calculate $\Delta U$. As can be seen in Fig. 1 and the convention used in Eqs. (1) and (2), a positive $\Delta E_{\text{VBO}}$ indicates the substrate valence band is higher in energy than the overlayer valence band, thus suitable for mobile hole confinement on the substrate side. As for the conduction band offset (CBO), the following relation is adopted:

$$\Delta E_{\text{CBO}} = \left(E_G^{\text{over}} - E_G^{\text{sub}}\right)_{\text{bulk}} - \Delta E_{\text{VBO}}, \quad (4)$$

where $E_G^{\text{over}}$ and $E_G^{\text{sub}}$ denote the bandgap in the respective material (see Fig. 1). Hence, $\Delta E_{\text{CBO}}$ is define positive when the conduction band of the overlayer is higher in energy, which also allows mobile electron accumulation near the interface in the substrate. In short, a positive (negative) offset indicates the barrier/well (well/barrier) alignment, respectively, at the overlayer/substrate interface for the corresponding type (p or n) of mobile carriers.

Another property of significance is the charge polarization characteristics. As described above, polarization at the interface can provide an effective means for doping the material for mobile carriers. The averaged macroscopic charge density is found from [31]

$$\bar{\rho}(z) = -\epsilon_0 \frac{d^2 \bar{V}(z)}{dz^2}, \quad (5)$$

where $\epsilon_0$ is the permittivity of vacuum and $\bar{V}(z)$ is the macroscopic average of the electric potential discussed above (more precisely, $-\bar{U}(z)/q$ with $q = 1.6 \times 10^{-19}$ C). The total 2D charge at the interface (per unit cross-sectional area) can then be obtained from the integration between two suitable endpoints on each side of the interface or, more conveniently, from the difference in the slope of $\bar{V}(z)$ at those two locations. Unlike some of the studies in the literature [32, 33], neither hydrogen termination nor dipolar correction is considered in this calculation to avoid contribution from any fictitious sources.

## III. RESULTS AND DISCUSSION

We consider three different interface orientations of diamond/c-BN, the different interface bonding configurations of those heterojunctions, and the two different substrate selections for a total of ten heterostructures. The first configuration is the structure with (110) interface which does not have alternate bonding configurations as C-B and C-N bonding are both present at the interface (see Fig. 2). Second is the (111) structure with C-B and C-N interface bonding configurations forming two separate heterojunctions. Third is the (100) configuration also with two separate systems for the C-B and the C-N bonding. For each geometry, calculations are performed with one material as the unaltered substrate and the other as the overlayer. The non-octet bonds of both the C-B and C-N (i.e., electron deficient and rich) imply charge polarization at the diamond/c-BN interface. In contrast, the (110) orientation should be non-polar with no net charge at the interface as both bonds exist in equal number; however, the imbalance of the bonds and the charge accumulation in the remaining systems can result in the possibility of reconstruction at the interface [26]. We consider only abrupt interfaces without reconstruction as it has been shown that they can be grown in appropriate environments (particularly, in the C-B case) [34].

Figure 3 shows the planar and macroscopic average potentials calculated in the (110) structure with diamond as the substrate. Due to the (near) absence of the charge polarization, the macroscopic average potential energy is indeed essentially flat in each of the bulk-like regions, enabling a straightforward estimation of $\Delta U$. Subsequent evaluation of $\Delta E_{\text{VBO}}$ from Eq. (1) gives 1.33 eV, which is within the range obtained by the earlier DFT studies reported in the literature (i.e., $\sim 0.71-1.42$ eV) [26, 35, 36]. The positive VBO indicates that the valence band in the c-BN overlayer is located lower than that of the diamond substrate. The corresponding offset for the conduction band is $-0.40$ eV using the experimental bandgaps of $E_G^{\text{dia}} = 5.47$ eV and $E_G^{\text{c-BN}} = 6.4$ eV [27]. In the case when c-BN is the substrate with the diamond layer on top, the VBO and CBO are found to be $-1.22$ eV and 0.29 eV, respectively. These numbers suggest that the valence maximum of c-BN consistently has a lower energy than that of diamond by about $1.2-1.3$ eV in the (110) configuration. Since the difference in the bandgaps ($\sim 0.93$ eV) is smaller than the estimated VBO, the conduction band minima in the two layers appear to be staggered in the same manner. Thus, the (110) diamond/c-BN structure has a type-II alignment independent on the choice of the substrate (i.e., diamond vs. c-BN). It is also interesting to note that the offsets in the two cases show only a small discrepancy in $\Delta E_{\text{VBO}}$ despite the different strain conditions (i.e., strain in c-BN vs. in diamond). This seems to suggest the relatively minor impact of the lattice mismatch on the band line-up compared to other factors in the diamond/c-BN combination [33].

When the calculation is applied to the (111) structure, two interface bonding configurations are considered explicitly. Figure 4 show the obtained results for both the C-B and C-N cases with c-BN as the overlayer. As expected, the macroscopic average of the potential energy exhibits more pronounced slopes from the polar nature of the bonding. For the estimation of $\Delta U$, the linear fits defined in the respective bulk-like regions (see the dashed lines) are extrapolated to the interface (see also Sec. 2). The obtained $\Delta E_{\text{VBO}}$'s are rela-tively small for these configurations, ranging from 0.23 eV (C-B) to 0.32 eV (C-N), while they are −0.21 eV (C-B) and −0.35 eV (C-N) for the diamond overlayer. Accordingly, our calculations consistently predict the valence band in the diamond layer to be about 0.2∼0.35 eV higher than that in c-BN independent of the substrate or bonding termination. Since this value for the VBO is substantially smaller than the difference in the bandgaps of the two constituent materials, a type-I alignment can be expected in the (111) structure with the CBO of ∼ 0.6−0.7 eV [19]. The results for the (100) systems are similar, for which $\Delta E_{\text{VBO}}$ of 0.20 eV (−0.30 eV) and 0.34 eV (−0.55 eV) are predicted for the C-B and C-N configurations with c-BN (diamond) as the overlayer, respectively. While the range for the VBOs is a bit broader (e.g., 0.2∼0.55 eV), it nevertheless indicates an alignment where the diamond conduction and valence bands are nested within the c-BN bandgap (i.e., type I). The fact that the numerical values of the offsets are rather consistent across differing substrates, bond terminations, and crystallographic orientations [perhaps the non-polar (110) structures excepted] as summarized in Tables I and II lends a degree of credence to the obtained results.

Comparison with other studies in the literature is not straightforward due to the disparate approaches to remedy the polar interfaces. For the (100) C-B configuration, a VBO of 0.79 eV was reported in Ref. [33], where both hydrogen passivation (of diamond and c-BN surfaces) and dipolar correction were employed in the DFT calculation. Another theoretical analysis adopting hydrogen termination in the diamond layer predicted $\Delta E_{\text{VBO}}$ of 0.045 eV and −0.587 eV (here, the diamond valence band being lower than c-BN) for the (111) C-B and C-N cases, respectively, where the partial density of states in the layers adjacent to the interface (rather than in the bulk regions) were used to estimate the offsets [32]. In addition, values of 2.0 eV and 0.13 eV were obtained in the (100) structures but with *mixed* C-B and C-N interfaces instead of the abrupt junctions. As with Refs. [26] and [35], these works used the average of the diamond and c-BN lattice constants for the supercells unlike the current investigation. While the reported numbers vary widely, it is reasonable to conclude that the valence bands of c-BN appears to have energies lower than those of diamond in the (111) and (100) structures as well.

Another interesting property to analyze in Fig. 4 is the presence of a peak (C-B) or a valley (C-N). A similar feature [i.e., a peak in the (100) C-B structure] was also observed in an earlier study [37]. The resulting change in the sign of the slope in the macroscopic average potential energy $\bar{U}$ clearly indicates the presence of sizable non-zero charges at/near the heterojunction. Figure 5 shows the planar average of the charge distribution [see also Eq. (5)] calculated in the (111) structure with diamond as the substrate (lower panels). As shown, the C-B and C-N junctions exhibit opposite polarization at the interface with net negative and positive values, respectively. The imbalance resulting from the non-octet bonding (i.e., electron deficient and electron rich, respectively) redistributes electrons toward or away from the interface. This point can be seen more clearly from the charge density difference plots provided in the upper panels, where the contributions from the isolated components are subtracted to highlight the electron transfer due to the bonding. Evidently, the C-B case attracts the electrons along the length of the bond, while the opposite occurs in the C-N configuration. This is also confirmed by the valley (−) and the peak (+) in $\rho$ right around the boundary between diamond and c-BN (marked by the vertical lines in the lower panels). The polarization region appears to be limited to the immediate vicinity of the heterointerface with the spatial spread of a lattice constant or so on each side. The resulting 2D net charge density in the atomically narrow region can be rather significant in both the (111) and (100) structures, as summarized in Table I. Similar results are also obtained in the cases where c-BN is used as the substrate (see Table II).

When mobile carriers become available in the system, the interface polarization discussed above is expected to draw those of opposite polarity toward the junction. Accordingly, a 2D electron or hole gas can be formed if the corresponding band offset can provide a proper confinement. Since both the (111) and (100) orientations are predicted to have a type-I alignment, it appears feasible to selectively achieve p-type and n-type channels in the diamond layer of these structures even without external doping (i.e., via C-B and C-N terminations, respectively). This mechanism of 2D carrier gas formation is analogous to the polarization doping commonly adopted in the nitride heterostructures. In the case of non-polar (110), a modulation doping scheme can be instead used to induce mobile carriers. In contrast to the scenario of p-type devices, the n-type counterpart in (110) (i.e., electrons confined in the c-BN potential well) may be more challenging to realize due to the relatively small $\Delta E_{\text{CBO}}$ (vs. the donor ionization energy in the diamond barrier). Needless to say, the modulation doping can also be applied to the (111) and (100) structures to enhance the 2D carrier density in the channel along with the polarization doping. Once formed, this 2D carrier gas is expected to exhibit superior transport properties due to the absence of external dopants and reduced scattering from confinement. Note that the small non-zero values for the (110) charge polarization in Tables I and II (e.g., mid $10^{10}$ to $10^{11}$ $q/\text{cm}^2$) are likely from the numerical averaging processes and may thus represent an error bar in our estimation.

## IV. SUMMARY

A theoretical analysis based on first-principles calculations is applied to examine diamond/c-BN heterostructures of different orientations and interface bonding for their feasibility in high-frequency, high-power device applications. The obtained band alignments show a strong dependence on the polar/non-polar orientation of the heterostructure, while the atomic configuration at the interface appears to have only a modest impact. In the non-polar (110) structure, the valence band maximum of the c-BN layer is predicted to be lower in energy than that of diamond with an offset of about 1.2−1.3 eV in a type-II alignment. In contrast, the diamond conduction and valence bands are likely to be nested within the c-BN bandgap, leading to a type-I configuration in both the (111) and (100) polar cases. The VBO and CBO are estimated to range ∼0.2−0.55 eV and ∼0.4−0.7 eV, respectively, sufficiently large for both mobile hole and electron confinement at the junction. These (111) and (100) structures are also shown to produce net charge polarization in an atomically narrow region at the interface. The electron-deficient and electron-rich nature of the C-B and C-N bonding induce charge redistribution leading to an essentially 2D sheet of negative and positive polarization on the order of $10^{12} - 10^{13}$ $q/\text{cm}^2$, respectively. The predicted band alignments suitable for carrier confinement as well as the possibility of the modulation and polarization doping make the diamond/c-BN heterostructures a promising candidate for high-performance electronic devices with a highly conductive 2D channel. Both p-type and n-type devices appear possible. Due to the near lattice match between the two materials, the results are found to be only weakly affected by the order of stacking as well (e.g., diamond/c-BN vs. c-BN/diamond).

| Diamond substrate Configuration | $\Delta E_{\text{VBO}}$ (eV) | $\Delta E_{\text{CBO}}$ (eV) | Interface 2D charge polarization ($q/\text{cm}^2$) |
|:---:|:---:|:---:|:---:|
| 110 | 1.33 | $-0.40$ | $+4 \times 10^{10}$ |
| 111 C-B | 0.23 | 0.70 | $-3 \times 10^{12}$ |
| 111 C-N | 0.32 | 0.61 | $+9 \times 10^{12}$ |
| 100 C-B | 0.20 | 0.73 | $-2 \times 10^{12}$ |
| 100 C-N | 0.34 | 0.59 | $+1 \times 10^{13}$ |

TABLE I: Valence and conduction band offsets ($\Delta E_{\text{VBO}}$ and $\Delta E_{\text{CBO}}$) and 2D charge polarization at the interface for the heterojunctions using diamond as the substrate. The bandgap difference of 0.93 eV ($= E_G^{\text{c-BN}} - E_G^{\text{dia}}$) is used to estimate $\Delta E_{\text{CBO}}$ from the calculated $\Delta E_{\text{VBO}}$. The positive numbers in $\Delta E_{\text{VBO}}$ and $\Delta E_{\text{CBO}}$ indicate the offset conditions favorable for 2D confining well formation in the substrate (in this case, diamond) for the respective band (i.e., p- and n-types, respectively). Accordingly, the $\langle 110 \rangle$ configuration is of type-II band alignment and the rest are type I. $q$ denotes the unit electronic charge of $1.6 \times 10^{-19}$ C.

| C-BN substrate Configuration | $\Delta E_{\text{VBO}}$ (eV) | $\Delta E_{\text{CBO}}$ (eV) | Interface 2D charge polarization ($q/\text{cm}^2$) |
|:---:|:---:|:---:|:---:|
| 110 | $-1.22$ | 0.29 | $+1 \times 10^{11}$ |
| 111 C-B | $-0.21$ | $-0.72$ | $-4 \times 10^{12}$ |
| 111 C-N | $-0.35$ | $-0.58$ | $+9 \times 10^{12}$ |
| 100 C-B | $-0.30$ | $-0.63$ | $-1 \times 10^{12}$ |
| 100 C-N | $-0.55$ | $-0.38$ | $+7 \times 10^{12}$ |

TABLE II: Valence and conduction band offsets ($\Delta E_{\text{VBO}}$ and $\Delta E_{\text{CBO}}$) and 2D charge polarization at the interface for the heterojunctions using c-BN as the substrate. As described in Table I, the positive numbers in $\Delta E_{\text{VBO}}$ and $\Delta E_{\text{CBO}}$ indicate the offset conditions favorable for 2D confining well formation in the substrate (in this case, c-BN) for the respective band. Just like the case described in Table I, the (111) and (100) structures are type I while the (110) orientation is type II.

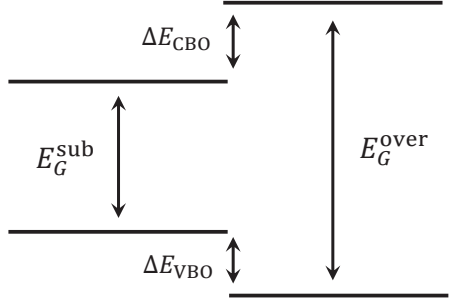

FIG. 1: Typical band alignment diagram. The substrate is depicted on the left and overlayer on the right. As shown, both $\Delta E_{\text{VBO}}$ and $\Delta E_{\text{CBO}}$ are defined "positive" since the corresponding mobile carriers (i.e., holes and electrons, respectively) can be confined on the substrate side of the heterostructure.

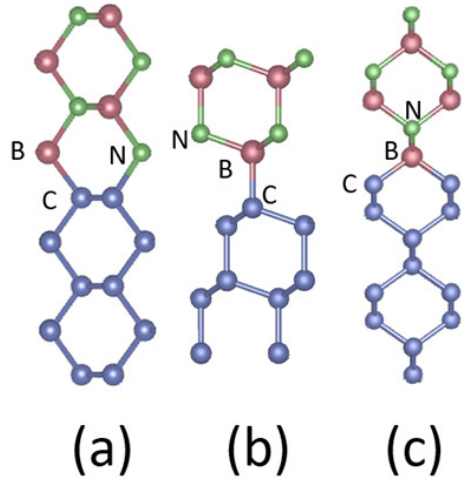

FIG. 2: Atomic configuration at the interface for three of the systems tested: (a) (110), (b) (111) with C-B bonding, and (c) (100) with C-B bonding. By switching the positions of B and N in (b) and (c), the corresponding structures for the C-N bonding can be obtained.

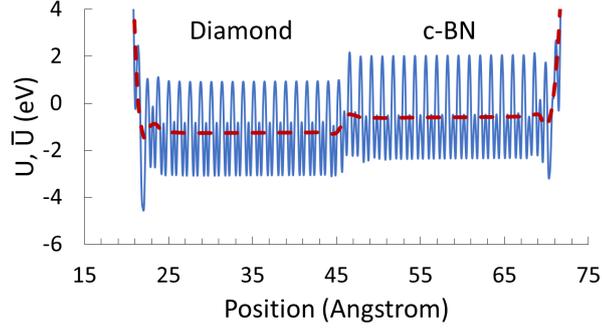

FIG. 3: Planar average (solid blue) and macroscopic planar average (dashed red) of the electrostatic potential energy for the (110) system (i.e., U and $\bar{U}$, respectively). The flat regions of the dashed line identify the bulk-like portions of the heterojunction.

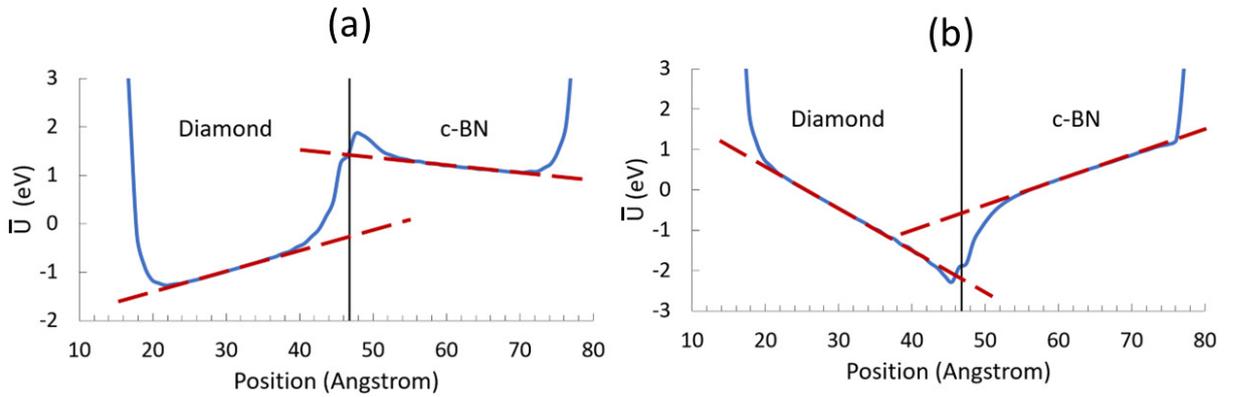

FIG. 4: Macroscopic averages $\bar{U}$ of the potential energy for (a) (111) C-B and (b) (111) C-N. The red dashed line is the linear fit for the bulk-like regions and the black vertical line denotes the boundary between the diamond and c-BN layers.

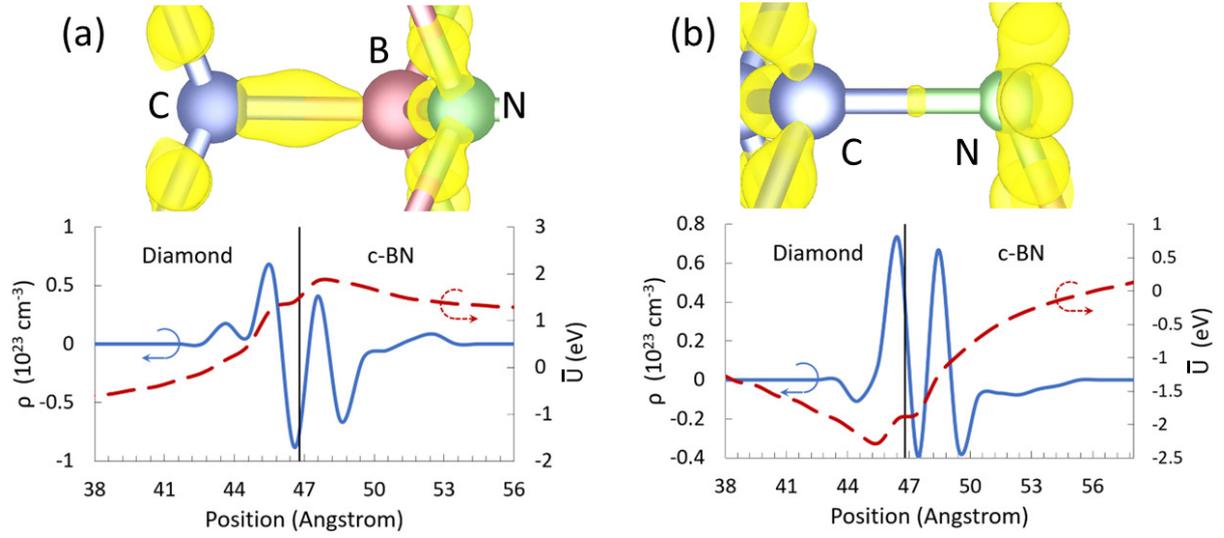

FIG. 5: 3D charge density difference (upper) and the macroscopic average of the charge density (lower, solid blue) along the direction perpendicular to the interface for the (a) (111) C-B and (b) (111) C-N bonding configurations. Isosurfaces (yellow; upper panel) show the 3D region where negative charge has accumulated (a) along the bond between C-B and (b) near the C and N atoms. The macroscopic average of the potential energy (dashed red) is included as an overlay. The black vertical line in the lower panel denotes the boundary between the diamond and c-BN layers.